\documentclass[np2,twoside,twocolumn,showpacs,floatfix]{revtex4}
\usepackage{amssymb}
\usepackage{amsmath}
\usepackage{url}
\usepackage{bm,multirow,color}
\usepackage{soul}
\usepackage{enumitem}

\definecolor{dkgreen}{rgb}{0,0.6,0}
\definecolor{gray}{rgb}{0.5,0.5,0.5}
\definecolor{mauve}{rgb}{0.58,0,0.82}

 \volumenumber{xx} \issuenumber{x}
\volumeyear{Month Year} 
\startpage{1}
\endpage{3}
 
\begin{document}

\title{Build Up of a Subject Classification System from Collective Intelligence}
 
\author{Jisung \surname{Yoon}}
\affiliation{Department of Industrial Management Engineering, Pohang University of Science and Technology, Pohang 37673, Korea}

\author{Jinhyuk \surname{Yun}}
\email{jinhyuk.yun@kisti.re.kr} 
\affiliation{Future Technology Analysis Center, Korea Institute of Science and Technology Information, Seoul 02456, Korea}

\author{Woo-Sung \surname{Jung}}
\email{wsjung@postech.ac.kr}
\affiliation{Department of Physics and Department of Industrial Management Engineering, Pohang University of Science and Technology, Pohang 37673, Korea\\
Asia Pacific Center for Theoretical Physics(APCTP), Pohang 37673, Korea}

\date[]{Submitted 10 December 2017}

\begin{abstract}
Systematized subject classification is essential  for funding  and  assessing  scientific projects. Conventionally, classification schemes  are founded  on the  empirical knowledge  of the  group  of experts; thus, the experts’  perspectives have influenced  the current systems  of scientific classification. Those systems  archived the current state-of-art in practice, yet the global effect of the accelerating scientific change  over  time  has  made  the  updating of the  classifications system  on a timely  basis  vertually impossible. To  overcome the aforementioned limitations,  we  propose  an  unbiased  classification scheme  that takes dvantage of collective  knowledge;  Wikipedia, an  Internet encyclopedia edited by millions  of users,  sets  a prompt classification in a collective  fashion.  We construct a Wikipedia network for scientific  disciplines  and  extract the  backbone of the  network. This  structure displays a landscape of science  and  technology that  is based  on a collective  intelligence and  that is 
more unbiased and  adaptable than conventional classifications.

\end{abstract}


\maketitle
\section{INTRODUCTIONS}\label{sec:introduction}
It is essential to establish the subject classifications as the instrument of information retrieval and decision making \cite{Sokal1974}. Naturally, humankind has produced various classification systems for the practical reasons, and the taxonomy for science and technology is one of the most demanded resources regarding the fact that science and technology is the complex of the knowledge composed of diverse correlated subjects. Since the assessment of research output highly depends on the current shape of classification system \cite{Waltman2011}, it necessitates constructing the unbiased classification system for the science and technology. Conventionally, a group of experts dedicate their expertise to these classification systems, e.g. ASJC (SCOPUS), MeSH (NCBI), PACS and PhySH (APS). For decades, these conventional classification scheme show practical results~\cite{Bornmann2010,Radicchi2011}, yet accelerating the changes of real world make it impossible to respond the current shape of science and technology timely due to the limitation of individuals' domain knowledge. Conventional classification schemes begin to disclose the limitations on flexibility, objectivity, and responsiveness. A new systematic way requested.\par
 
On the opposite site, scholars attempted to describe the shape of science and technology by mapping of paper citation data to overcome the aforementioned limitations. They proposed the two-level hierarchical system of natural science, social science and arts\&humanities with the three-step iterative process: cognitive-, pragmatic-, and scientometric- approaches \cite{glanzel2003}. Nested maps of science and technology with $14$ factors, $172$ categories, and $6\,164$ journals were also proposed \cite{Leydes2009}. Inspired by the similarity of the network community structure and scientific classification, community detection algorithm that classifies the \textit{nodes} by cluster are also applied. As the illustrative example, scholars obtained a cognitive map of astrophysics using topic affinity network with the \textit{Infomap} algorithm \cite{velden2017,Rosvall2008}. Also, the publication-level classification of research publications was also detected with the modularity optimization and compared with the pre-defined external classification \cite{Palchykov2016}. However, stressing the fact that the overall journal review process accompanies long delay on publication, citations also accompanies time lags. Accordingly, citation data of paper also cannot trace fast-changing landscape of contemporary science and technology.\par

Here, on the basis of collective intelligence data, we attempt the systematic and quantitative extraction of the structure of science and technology in the society. Collective intelligence is the knowledge that emerges from the systems of multiple agents, which evolves with the competition and cooperation of many entities in systems. Naturally, it has high receptivity for the social change; we thus choose to take advantage of such collective intelligence. Accordingly, we supposed that an extensive encyclopedic collective intelligence data produced on the web might be suitable for such analysis. Wikipedia is free on-line encyclopedia editing by volunteers who edit and debate the contents in real time. As the representative example of collective intelligence, Wikipedia data has been widely investigated. The dynamics and pattern of modification in Wikipedia were investigated, and clear evidence of bursty activity patterns is discovered\cite{Yasseri2012a,Yasseri2012b}. Oligarchy structure in editing behavior also reported with the mechanistic model for intellectual interchanges \cite{Yun2016}. Taxonomy of Wikipedia was extracted by integrating WordNet and Wikitaxonomy \cite{Ponzerro2009}.Also, the category structure of Wikipedia was decoded as the concepts and discovered the relations between concepts according to the network structure \cite{Nastase}. They found that the titles are an abundant collection of human knowledge. Wikipedia reflects the landscape of the contemporary science and technology without any delay.\par

In this paper, we construct the classification scheme for science and technology with extracting the backbone of the network on the advantage of collective intelligence. The rest of the paper is organized as follows. In Section \ref{sec:network_construction}, we describe the Wikipedia meta data that use in our study. In section \ref{sec:methods}, we introduce the method to extract the backbone of science and technology. We demonstrate the extracted classification scheme can be verified with the raw network and the All Science Journal Classification (ASJC) that the new backbone does not ruin the essential properties of both systems in Section \ref{sec:result}. Finally, we present our conclusion and discussion in Section \ref{sec:conclusion}.
 
\section{DATA AND NETWORK CONSTRUNCTION} \label{sec:network_construction}
For the analysis, we use the September $09$, $2017$ dump of the metadata and category link data of Korean Wikipedia. For each metadata of the article, there is \texttt{name-space} element, which indicates the role of the article in $34$ types. Specifically, we used the articles with namespace $0$ and $14$, which are the contents page and category page, respectively. Category-link data including the reference relations from articles pointing to the category also used to construct the network as the link for science and technology classification scheme.\par 

We construct a directed and unweighted network from the data above. We assign a node as a category or a page. A directed link connects if node A refers node B. As the illustrative example, if \texttt{complex system} refers \texttt{statistical mechanics}, we then assign directed link from \texttt{complex system} to \texttt{statistical mechanics}. Because we consider both the articles and categories as the node distinguished from the name, there is a possibility of homonyms from the different name-space, e.g. \texttt{Category:Natural science} and \texttt{Article:Natural science}; we thus merge multiple nodes sharing the names into a single node. The category network is to describe the entire interrelationship between the articles in Wikipedia that is essentially accompanying the surplus nodes, so the filtering is required to construct the classification scheme of science and technology. We thus use the subnetwork of nodes that can be reachable to \texttt{Category:scientific discipline (분류:분야별 과학 in Korean)}. As a result, the subnetwork is composed of $568\,676$ nodes and $2\,251\,667$ links in total.\par

\begin{figure}[!ht]
\includegraphics[width=8cm]{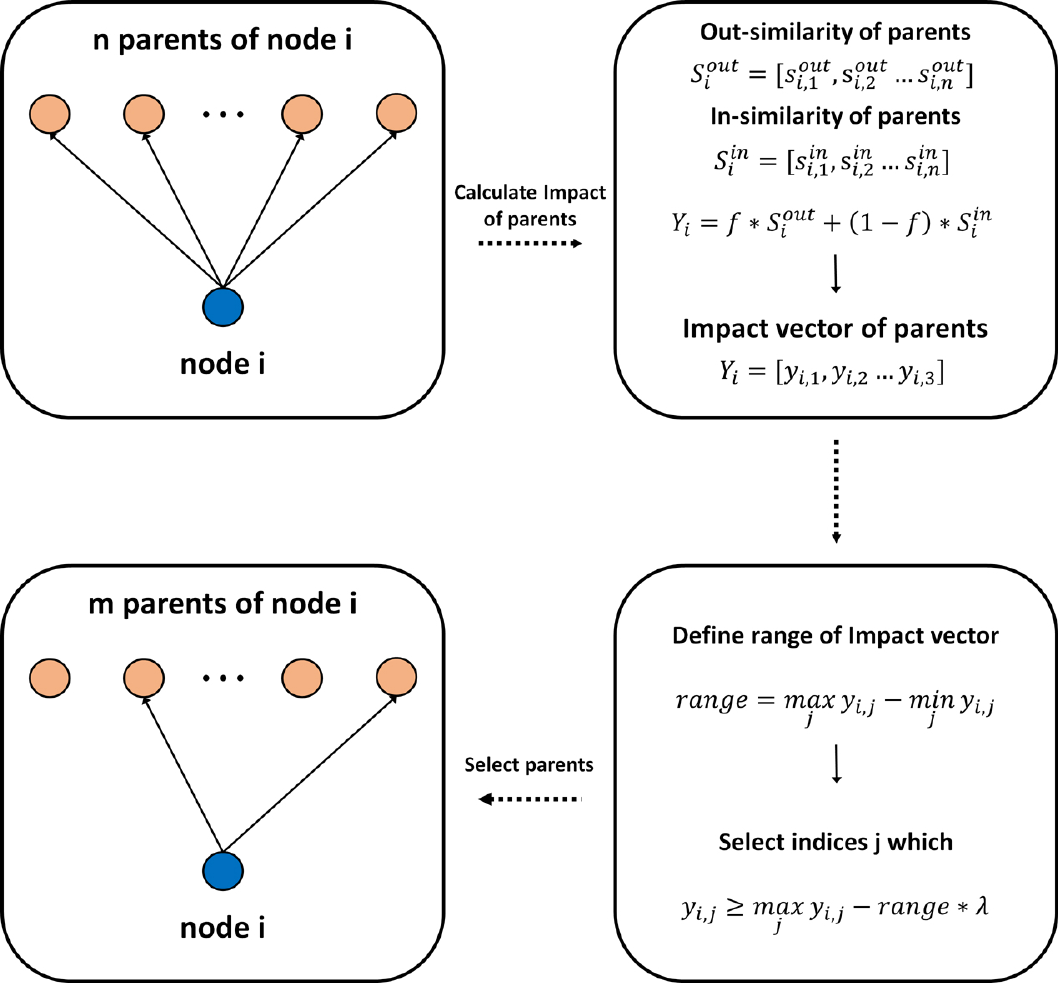}
\caption{(Color online) The procedure of the reduction process.}
\label{reduction_method}
\end{figure}

\section{EXTRACTING THE BACKBONE OF THE NETWORK}\label{sec:methods}
Although we already remove the inessential nodes for the scientific classification in Section \ref{sec:network_construction}, the network is too dense to utilize as the scientific classification scheme; uncovering a hierarchical structure is needed. We extract the backbone of the network with following processes. i) \textit{Pruning Process} prunes insignificant links using shortest path information. ii) \textit{Reduction Process} reduces the network using local structural information of the network.

\subsection{Pruning Process}\noindent
As the first step, we remove redundant links from the shortest path information. We compute entire shortest paths of all nodes to the root node (\texttt{Scientific discipline}) and check whether a certain link is used for a certain shortest paths or not. If the link does not belong to the any of shortest paths, we remove the link. During the pruning process, $\sim61\%$ redundant links are eliminated. Specifically, only $871\,766$ links remain in the pruned network out of $2\,241\,667$ links (see Table~\ref{table1}).

\noindent\subsection{Reduction Process}
For the next step, we now employ the backbone extracting method that uses the information of common neighbors of two connected nodes \cite{Gualdi2011, Valverde2015}. We calculate impact vectors $Y_{i} = [y_{i,1},y_{i,2},...,y_{i,n}]$ for $n$ parents of the given node. The elements of impact vector displays the influence of a certain parent node to the give node, which is computed by linear combination of an out-similarity vector $S_i^{out} = [s_{i,1}^{out},s_{i,2}^{out},...,s_{i,n}^{out}]$ and an in-similarity vector $S_i^{in} = [s_{i,1}^{in},s_{i,2}^{in},...,s_{i,n}^{in}]$, which can be written as
 \begin{equation}
 	Y_i = f * S_i^{out} + (1-f) * S_i^{in},
 \end{equation}
where $f$ is a control parameter that determines the ratio of reliance between ancestors and descendants. This can be chosen in real number between $0$ and $1$, yet we choose the most basic value of $f=1/2$. Here, the similarity $S_{i,j}^{out}, S_{i,j}^{in}$ is defined as 
  \begin{align}
    S_{i,j}^{out} = a_{ij}\sum_{t\neq i}\frac{1}{k_t^{out}} \sum_{l}\frac{a_{il}a_{tl}}{k_l^{in}},\\
 	S_{i,j}^{in} = a_{ij}\sum_{t\neq i}\frac{1}{k_t^{in}} \sum_{l}\frac{a_{li}a_{lt}}{k_l^{out}},
 \end{align}
 where $a_{ij}$ is $(i,j)^{th}$ a element of the network's adjacency matrix which is $1$ if there exists a link from $i$ to $j$, $0$ otherwise, and $k_t^{out}, k_t^{in}$ is out-degree and in-degree of the node $t$, respectively. 
We then define a set of peer nodes $\pi_j$, which are children of node $j$ except node $i$ to calculate $S_{i,j}^{out}$ ($S_{i,j}^{in}$).
For out(in)-similarity, the similarity is equivalent as the summation of the probability that random walker starting at peer node $l$ reaches the destination node $i$ across the common ancestor(descendant) node $t$ in two step. For the out-similarity, the walker moves along the direction for the first step, whereas the walker traces back as the reverse direction of the link for the second step. The order is reverse in case of in-similarity. \par

Once we compute the impact vector, we reduce the networks by removing the links with small impacts. Specifically, we assign the reduction cutoff $\lambda \in [0, 1]$ (real numbers between 0 and 1) that determine how much information of the pruned network preserved in the reduced network. We then delete links to the nodes whose impact value $y_{i,j} < \max_j(y_{i,j}) - \lambda\left[\max_j(y_{i,j}) - \min_j(y_{i,j})\right]$ (see Fig.~\ref{reduction_method}). As the extreme cases, all nodes with maximum values are selected for $\lambda = 0$ and the pruned network without reduction is retained for $\lambda = 1$. One should note that overall structure of the reduced network varies with the reduction cutoff $\lambda$; thus, finding adequate $\lambda$ is crucial. Accordingly, we set three criteria for appropriate $\lambda$ and reduce the network until the following conditions are satisfied:

\begin{enumerate}[noitemsep]
    \item All nodes in the original network is remained in the reduced network.
 	\item The reduced network is in a single component.
    \item The root node is reachable from all nodes in the reduced network (at least one path).
\end{enumerate}
  
\begin{table}[!ht]
  \begin{center}
    \caption{Size of the extracted backbones of the network: i) the number of the nodes in the largest component (N), ii) the number of the links (M), iii) link-node ratio (r), and iv) the percentage of remained links compare with original network($\%_M$) for different values of reduction cutoff $\lambda$.}   
    \label{table1}
    \begin{tabular}{ccccc} 
      \hline \hline
      \textbf{$\lambda$} & \textbf{$N$}& \textbf{$M$}&\textbf{$r$}&\textbf{$\%_M$}\\
      \hline
      Original & $568\,876$ & $2\,241\,667$ & $3.941$&$100\%$\\ 
      Pruned & $568\,876$ & $871\,766$ & $1.532$&$38.89\%$\\  
      0.1 & $568\,876$ & $577\,497$ & $1.015$&$26.76\%$\\ 
      0.01 & $568\,876$ & $571\,787$ & $1.005$&$25.51\%$\\ 
      0 & $568\,876$ & $571\,110$ & $1.003$&$25.48\%$\\ 
      \hline \hline
    \end{tabular}
  \end{center}
\end{table}

\begin{figure}[!ht]
\centering
\includegraphics[width=8cm]{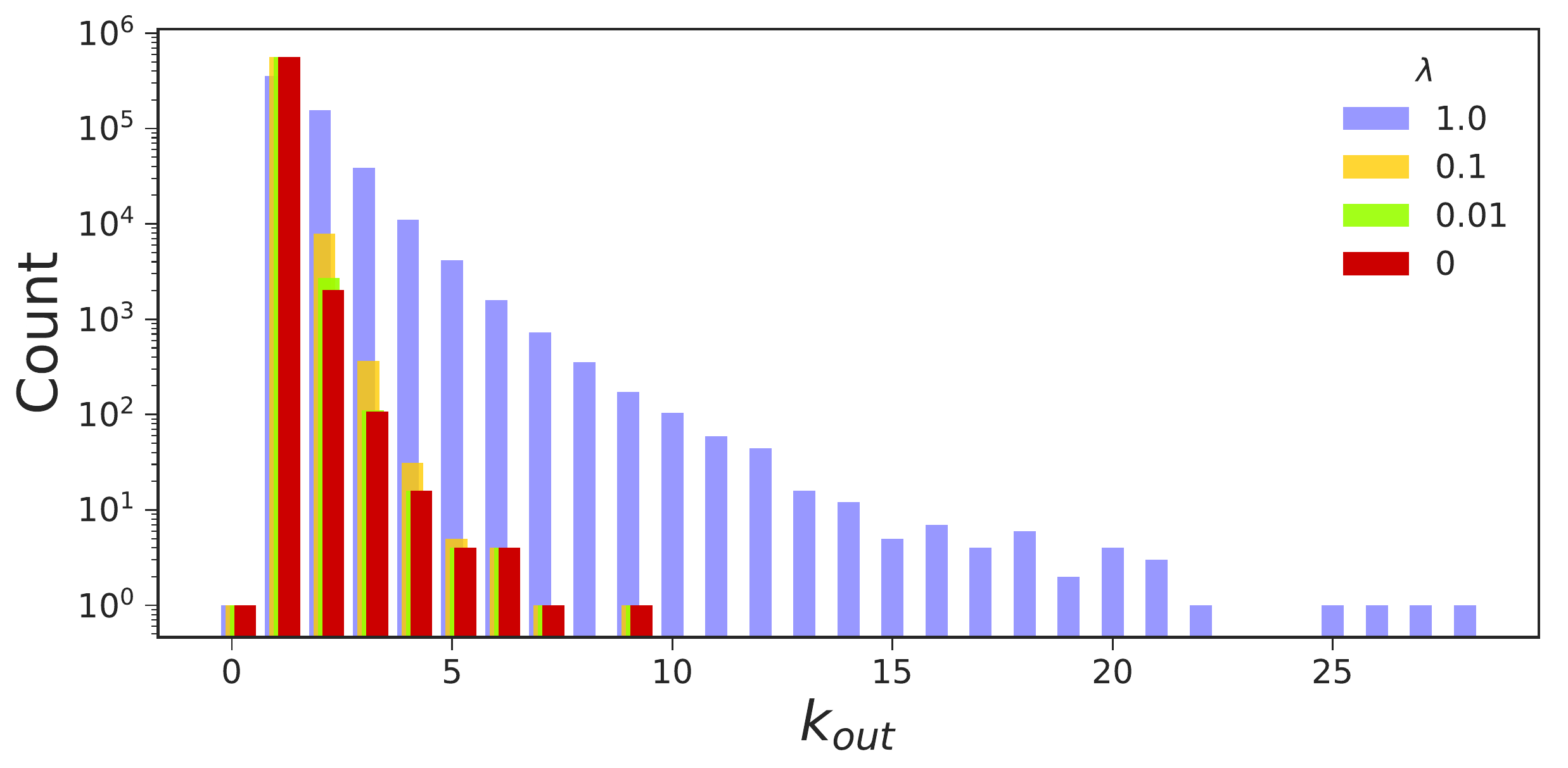}
\caption{(Color online) The number of distinct nodes as the function of out-degree $k_{out}$. Each color corresponds to the reduction cutoff $\lambda$, which is control parameter.} 
\label{out_degree}
\end{figure}

\section{RESULTS}\label{sec:result}
In Section~\ref{sec:methods}, we propose the reduction method to extract category structure of science and technology in Korean Wikipedia. A logical step forward is to assess the possibility to use the reduced network as the alternative classification scheme for science and technology. In this section, we inspect the suitableness of the reduced network as the taxonomy. \par

\subsection{Properties of reduced network}
We discover that the three criteria set in Section~\ref{sec:methods} are not violated for our reduced network. Even though we choose the smallest $\lambda$, the number of nodes in the largest component maintained, whereas the number of links consistently decreases with smaller $\lambda$ (criteria 1 and 2; see Table~\ref{table1}). Also, all nodes are reachable to the root node for all range of $\lambda$ (criteria 3). The network is becoming sparser when reduction cutoff $\lambda$ is closer to $0$. Only $25.48\%$ of the links are remained in the extracted backbone with $\lambda = 0$ in compare to the original network. In addition, We observe the wide ranged out-degree distribution for larger value of $\lambda$ (see Fig.~\ref{out_degree}). For example, there is a node has $28$ parent nodes for $\lambda = 1$. Because the smallest $\lambda$ ($\sim 0$) also meet all three criteria, we set the smallest reduction parameter $\lambda=0$ for the following analysis. \par

\begin{figure}[!ht]
\centering
\includegraphics[width=8cm]{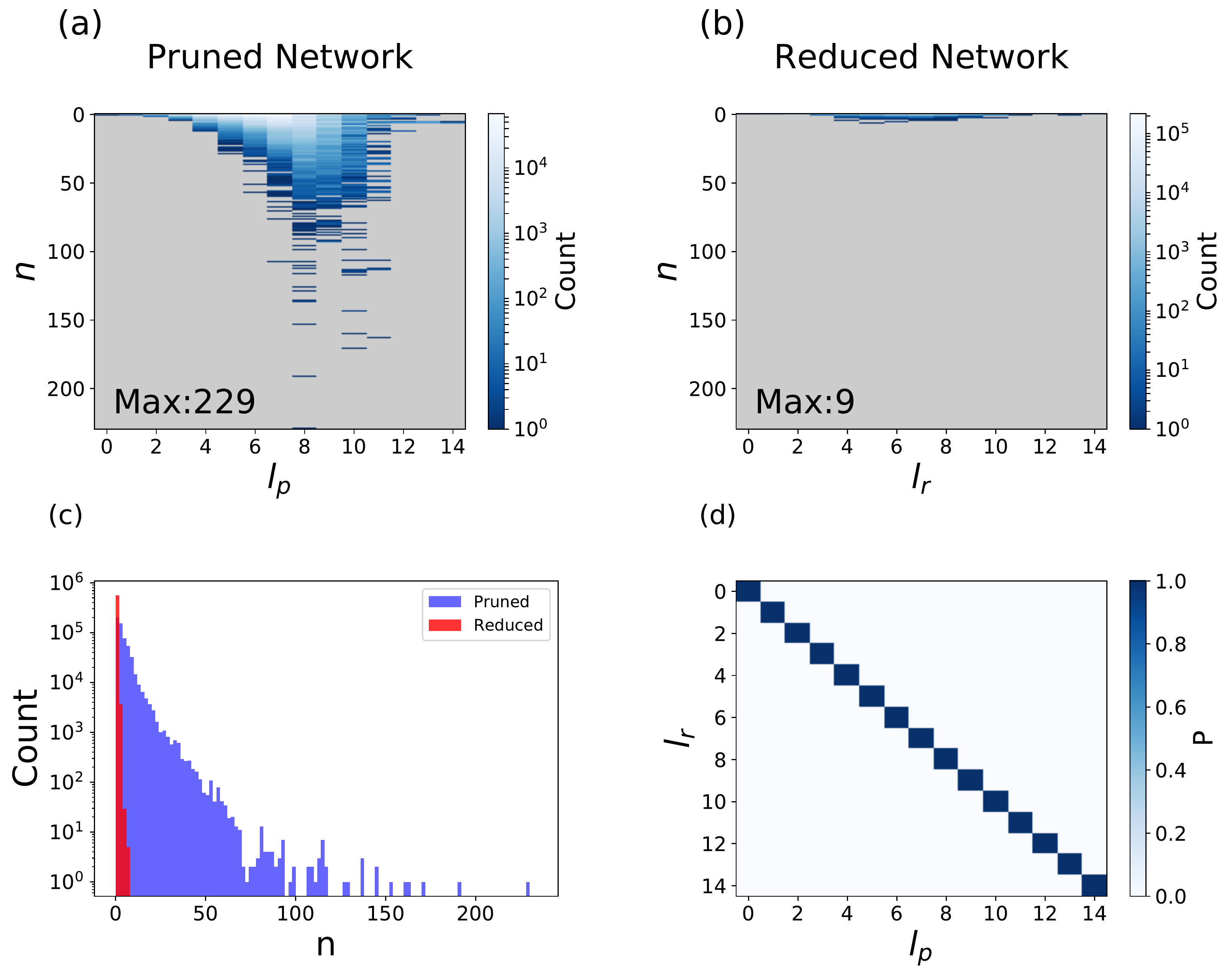}
\caption{(Color online) Properties of reduced network. (a) Numbers of alternative shortest paths $n$ for a given node as the function of the shortest path length of the pruned network $l_p$. (b) Numbers of alternative shortest paths $n$ for a given node as the function of the shortest path length of the reduced network network $l_o$ with reduction cutoff $lambda = 0$. For (a) and (b), the color indicates the number of nodes for a certain $n$ and shortest path [$l_p$ for (a) and $l_r$ for (b)] (c) Distributions of the number of the shortest paths of given node in the pruned network and the reduced network. (d) Correlation between between $l_p$ and $l_r$. For (d) the color denotes the number of nodes of a certain pair of $l_p$ and $l_r$ }
\label{main_result}
\end{figure} 

The next question is whether nodes in the reduced network can be sorted into clusters sharing common topics with hierarchy. In other words, if the network function as the classification scheme, the nodes should have an acceptable number of shortest paths to the root node, otherwise the network can be clustered into a single group with the complex interrelations. Indeed, nodes in the pruned network have many alternative paths, e.g. a node with $229$ shortest paths to root nodes [see Fig.~\ref{main_result}(a)]. In that sense, the pruned network is improper to use as the classification scheme. In contrast, nodes in the reduced network have an adequate number of the shortest paths compared to the pruned network [see Fig.~\ref{main_result}(b) and (c)]. \par

\begin{figure}[!ht]
\centering
\includegraphics[width=8cm]{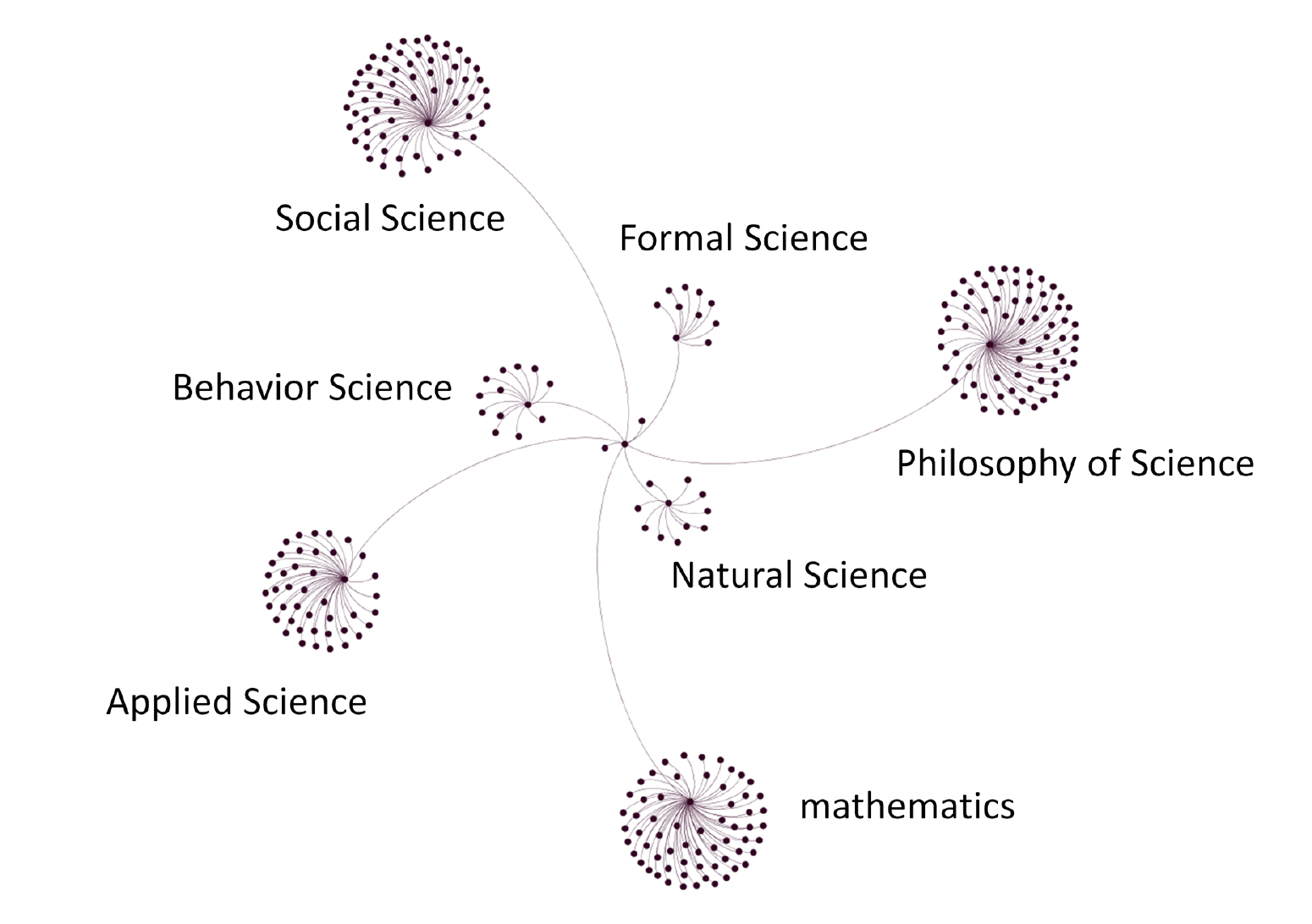}
\caption{(Color online) Illustrative example of the reduced network. The graph is drawn up to second level that is defined as the shortest path length from the root node. Each label corresponds to the name of the first level node of the group of the nodes.}
\label{level2}
\end{figure}

\begin{figure}[!ht]
\centering
\includegraphics[width=8cm]{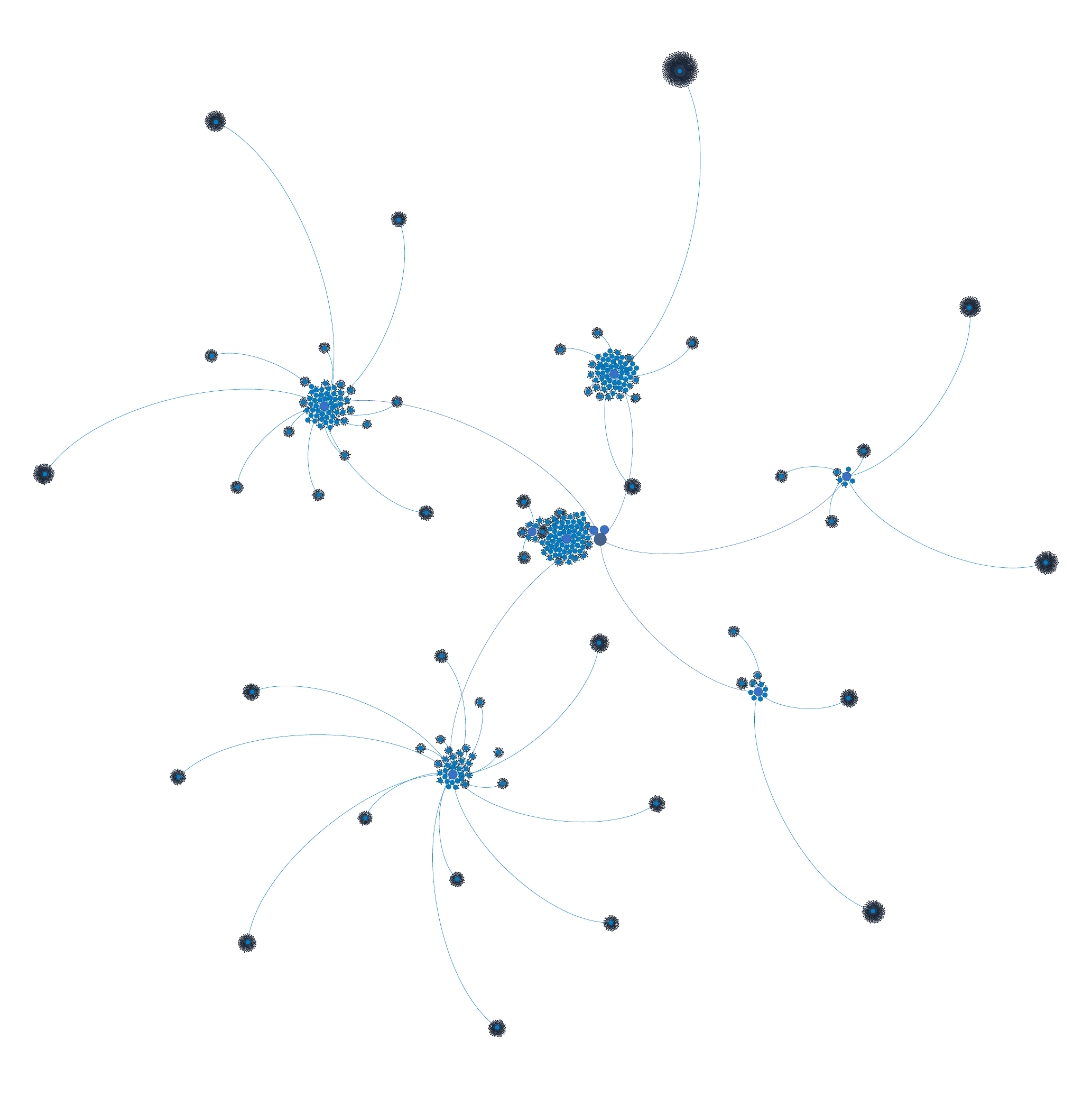}
\caption{(Color online) Illustrative example of the reduced network drawn up to third level. The level is defined as the shortest path length from the root node.}
\label{level3}
\end{figure}

\begin{figure}[!ht]
\centering
\includegraphics[width=8cm]{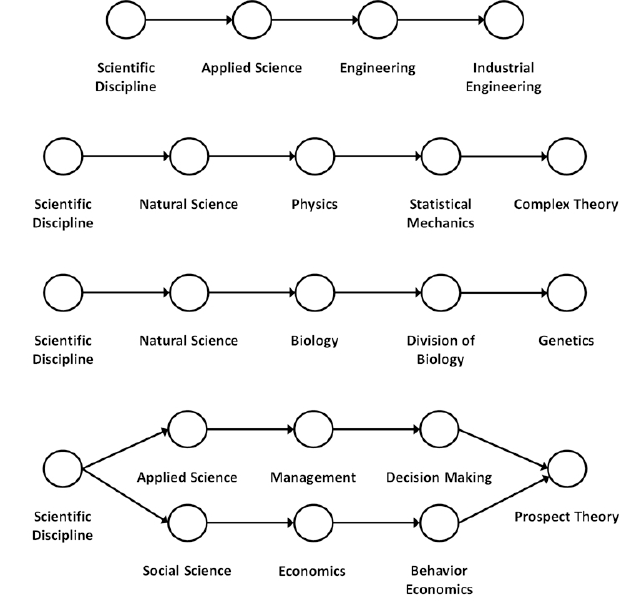}
\caption{(Color online) The sample paths of the constructed classification scheme with collective intelligence. In this figure, the arrow is drawn in the reverse direction of the actual link.}
\label{sample_path}
\end{figure}

Moreover, our filtering method does not change the shortest path length. The influence of the filtering is strictly limited to the cutting redundant information without ruining the hierarchal levels of the nodes. Therefore, the processed network can be used as the classification scheme for science and technology. The illustrative example is shown in Figs. \ref{level2} and \ref{level3}, which is the visualization of the network up to the second level and third level, respectively. Both visualizations display the hierarchal structure as well: seven major areas of science and technology appear in the first level, subareas is attached like branches of a tree. Interestingly, our classification system has a multi-routed path to the discipline that is absent in conventional classification system (Fig.~\ref{sample_path}), which give the better description for the disciplines from interdisciplinary sources. \par

\def\msun{M_\odot}

\subsection{Verification with a conventional classifications}
Although reduced Wikipedia classification scheme is well-defined to account the structure of science and technology described in Wikipedia, the new classification scheme should be corresponding to the conventional system, e.g. All Science Journal Classification (ASJC), at least to some degree for practical use. For comparison, we identified the equivalence ASJC entities for each nodes in the reduced Wikipedia category network. The number of entities in ASJC is considerably smaller than those of Wikipedia category network. Naturally, entities in ASJC can be matched with the multiple nodes of the reduced category network. For example, \texttt{History and philosophy of science} in ASJC matched to \texttt{History of science} and \texttt{Philosophy of science} in the reduced category network. Additionally, only 238 ASJC subjects out of 329 subjects are matched with the extracted classification scheme because we neglect the fragmentary matched subjects to minimize the ambiguity. \par

\begin{figure}[!ht]
\centering
\includegraphics[width=8cm]{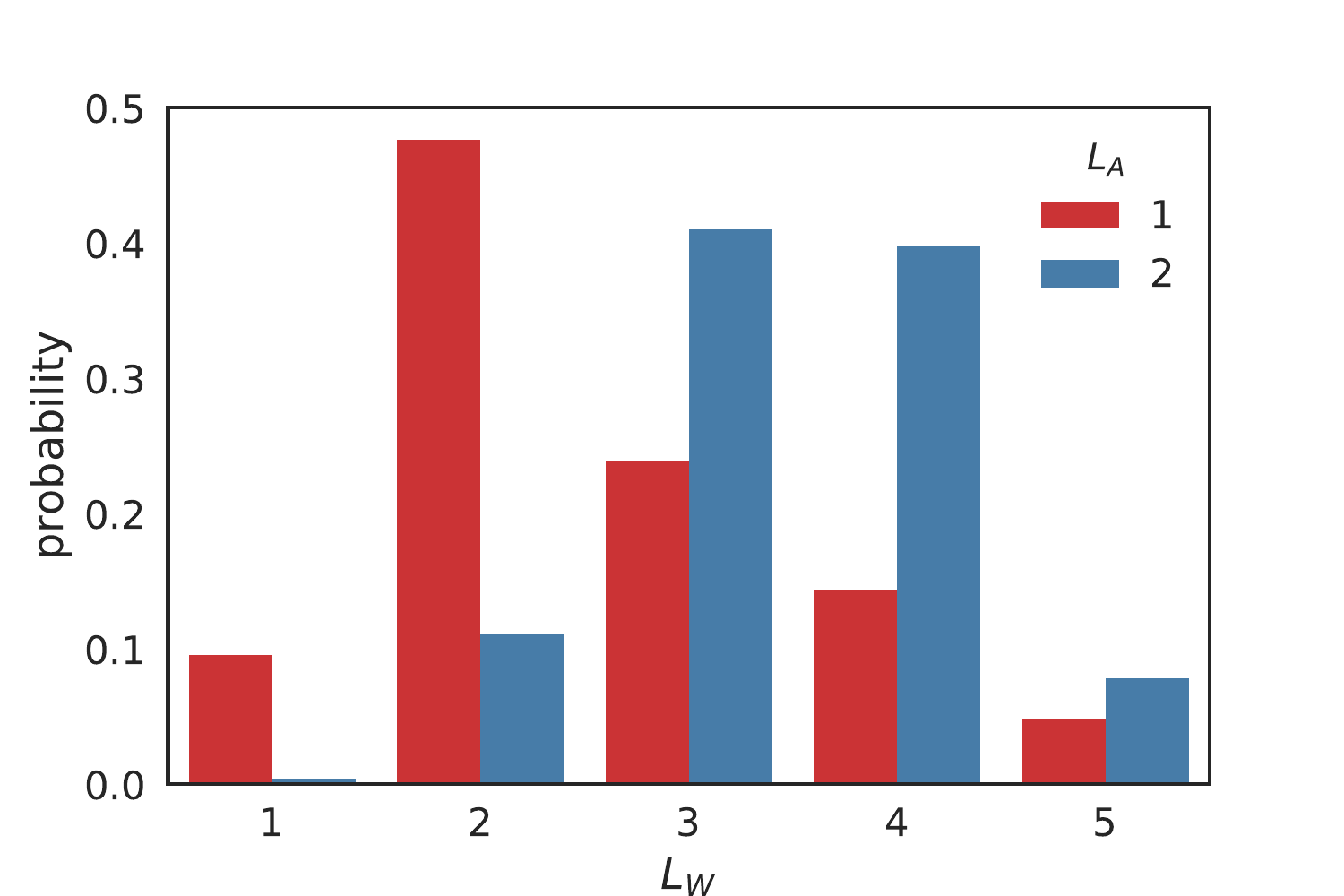}
\caption{(Color online) The distributions for the level of reduced Wikipedia category network for a given ASJC level. $L_w$ denotes the level of the extracted classification scheme and $L_A$ denotes the level of the ASJC.}
\label{level_compare}
\end{figure}

For verification, we define the levels as $1$ for the subject areas (large categories) and $2$ for the subject categories (small categories) for ASJC classification scheme, whereas the shortest path length from the root node is used as the level for the reduced Wikipedia classification scheme. We then perform the correlation analysis between the levels of entries for two classification schemes. Kendall's $\tau$ rank correlation measures the ordinal association between two sequences that range from 0 for the minimal correlation to 1 as the maximal correlation. We used the modified version of Kendall's $\tau$ rank correlation ($\tau_b$), to account the multiple ties essentially accompanied in the level \cite{kandell1955}. We observe the significant positive correlation between two levels ($\tau_b \simeq 0.46$) reflecting the acceptability of the conventional subject classifications. Fig.~\ref{level_compare} shows the distributions of the reduced Wikipedia category network for the different levels for matched ASJC categories. Two distributions clearly fall into the different regime. Level $1$ nodes mainly occupy higher levels (from $1$ to $3$) and level $2$ nodes located in the lower level (from $3$ to $4$). Interestingly, both ASJC subject categories and areas do not match with the Wikipedia categories with the level lower than 6, indicating our new classification is more segmentalized than the conventional classification scheme. KL-divergences that measure the distance between two distributions \cite{kullback1951} are $1.01$(first level$\to$second level) and $0.711$(second level$\to$first level). \par

\begin{figure}[!ht]
\centering
\includegraphics[width=8cm]{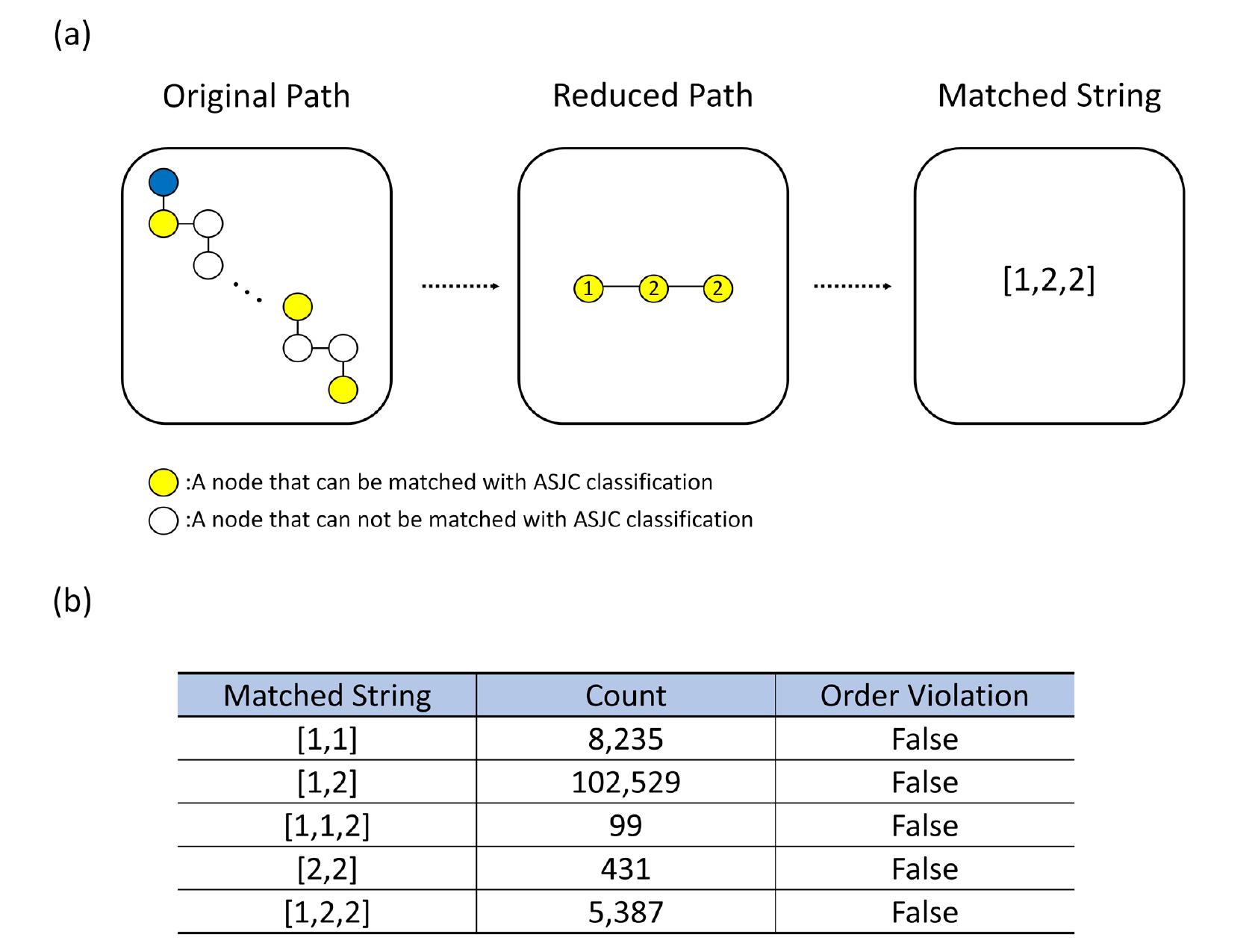}
\caption{(Color online) Schematic diagrams and result of the validation method.}
\label{validation_method}
\end{figure}

It also should be noted that the order of level should not be reversed in the new classification scheme for the consistency, between two classification scheme. Accordingly, we perform the analysis on the order of matched ASJC levels by generating a minimal ASJC level sequence of reduced Wikipedia category network (see Fig.~\ref{validation_method}(a). First, we select all shortest paths to root node in the extracted classification scheme. In these paths, there are some nodes that can be matched with ASJC or not. We then drop the nodes that are not matched with ASJC to extract the minimal matched path. Finally, the minimal ASJC level sequence is generated by replacing the node to the level of the matched ASJC areas in the reduce path. For the sequence, we consider order is violated if the lower level, i.e. 2, appears before the higher level, i.e. 1. As expected, all existing sequences do not violate the ordinal sequences Fig.~\ref{validation_method}(b). Additionally, the strings that have consecutive sequences observed. This finding may imply that reduced Wikipedia category network is more segmented than the coarse classification in ASJC classification scheme, which is coherent with the level distribution itself (see Fig.~\ref{level_compare}). \par

By incorporating the above evidences, reduced Wikipedia category network has clear consistency against the conventional classification scheme, i.e. ASJC. Moreover, considering the detailed subdivisions in our new classification system, it has clear advantage to describe contemporary science and technology compared to the conventional systems.\par

\section{CONCLUSION}\label{sec:conclusion}
In this study, we extract the Wikipedia category network taking advantage of collective intelligence. A new classification system has clear advantages: it has higher receptiveness and responsiveness on the change of science and technology. We propose the guidelines for deriving the reduced network as classification scheme with criteria, and verify our method does not violate any criteria in all cases. In addition to the flexibility, the new classification scheme gives the more detailed description of the scientific structure than conventional ones (comparing fourteen levels depth of our classification with two levels of ASJC classifications). We suggest that in-depth analysis into interrelation between scientific research and Wikipedia article may be warranted to enhance the impact of our approach. Also, extending our analysis to other language editions other than Korean Wikipedia and refining the presented reduction method are left for the further study. We hope that our approach sheds light on establishing the unbiased real-time classification system of the science and technology in the future. \par

\section*{ACKNOWLEDGEMENTS}
This work was supported by the POSTECH Basic Science Research Institute Grant (Jisung Yoon and Woo-Sung Jung), and the National Research Foundation of Korea Grant funded by the Korean Government through Grant No. NRF-2017R1E1A1A03070975 (Jinhyuk Yun). The funders had no role in study design, data collection and analysis, decision to publish, or preparation of the manuscript.

\end{document}